\documentclass[conference]{IEEEtran}
\IEEEoverridecommandlockouts
\usepackage{cite}
\usepackage{amsmath,amssymb,amsfonts}
\usepackage{algorithmic}
\usepackage{graphicx}
\usepackage{textcomp}
\usepackage{xcolor}
\usepackage{tikz,float}
  \newlength\fheight
\newlength\fwidth
\usepackage{tkz-euclide}
\def\BibTeX{{\rm B\kern-.05em{\sc i\kern-.025em b}\kern-.08em
    T\kern-.1667em\lower.7ex\hbox{E}\kern-.125emX}}
    \usepackage{pgfplots} 
\usepackage{pgfgantt}
\usepackage{pdflscape}
\pgfplotsset{compat=newest} 
\pgfplotsset{plot coordinates/math parser=false}
\pgfplotsset{every  tick/.style={black,},ylabel style={font=\tiny},xlabel style={font=\tiny},tick label style={font=\tiny},legend style= {font=\scriptsize},
minor x tick num=1,minor y tick num=1,xminorticks=true,yminorticks=true,}    
\begin{document}

\title{Zero-Forcing Max-Power Beamforming for Hybrid mmWave Full-Duplex MIMO Systems
}

\author{\IEEEauthorblockN{Elyes Balti}
\IEEEauthorblockA{\textit{Wireless Networking and Communications Group} \\
\textit{The University of Texas at Austin}\\
Austin, TX 78712, USA \\
ebalti@utexas.edu}
\and
\IEEEauthorblockN{Neji Mensi}
\IEEEauthorblockA{\textit{Department of Electrical Engineering and Computer Science} \\
\textit{Howard University}\\
Washington, DC 20059, USA \\
neji.mensi@bison.howard.edu}
}

\maketitle

\begin{abstract}
Full-duplex (FD) systems gained enormous attention because of the potential to double the spectral efficiency. In the context of 5G technology, FD systems operating at millimeter-wave (mmWave) frequencies become one of the most promising solutions to further increase the spectral efficiency and reduce the latency. However, such systems are vulnerable to the self-interference (SI) that significantly degrades the performance. To overcome this shortcoming, analog-only beamforming techniques have been developed to mitigate the SI. Because of the huge power consumption, systems operating at mmWave frequencies beamform the power by only tuning the phase shifters while maintaining constant amplitudes. Such a hardware constraint, known as the constant amplitude (CA) constraint, severely limits the system performance. In this work, we propose a digital and analog hybrid beamforming design that completely eliminates the SI while substantially minimizing the losses imposed by the CA constraint. Further, we develop a fully-digital beamforming design and derive the upper bound for the spectral efficiency as benchmarking tools to quantify the losses of our proposed hybrid design.
\end{abstract}


\section{Introduction}
With the exponential growth of the number of mobile users, cellular network systems operating in microwave and sub-6-GHz bands reached a bottleneck mainly because of the limited available bandwidth. The capacity of the conventional network becomes limited and is unable to support the current and future generation network which is known to be ultra densified. Such conventional systems also suffer from long latency during the uplink and downlink resource allocations and the initial channel access. Fortunately, millimeter-wave (mmWave) technologies have been emerging as a promising solution to address the aforementioned limitations. MmWave technologies are characterized by huge available bandwidth that covers from 28 to 300 GHz. In addition, they have been introduced to densify the cellular system, increase the spectral efficiency, reduce the latency, and improve the network scalability. Furthermore, mmWave-based commercial standards and products have been developed, such as IEEE 802.11 ad wireless gigabit alliance (WiGig), 5G modem, 5G new radio (NR), and mmWave prototype \cite{rheath,5gnr}.

To further improve the spectral efficiency, full-duplex (FD) systems have been introduced in the context of mmWave technologies. It has been shown that FD systems have the potential to increase and double the spectral efficiency compared to the conventional half-duplex systems \cite{e1,e2,e3,e4,e5,e6,e7}. However, FD systems suffer from the self-interference (SI), since the transceiver transmits and receives in the same frequency band. Without an appropriate SI cancellation, FD systems experience severe losses that significantly reduce the spectral efficiency. 

Different techniques for SI cancellation have been introduced, such as antenna separation, isolation, polarization, directional antennas, or antenna placement to create null space at the receiving (RX) arrays \cite{ant}. These methods depend on the physical architecture of the system. For example, antenna separation can achieve significant SI reduction when the transmitting (TX) and receiving arrays are not collocated. Such technique is not recommended for small devices since the TX and RX arrays cannot be sufficiently separated. Moreover, SI cancellation can be realized by analog and digital cancelling circuits. For example, the QHx220 chip requires the knowledge of the transmitted SI, tunes its phases and magnitudes to match the received SI signal, and subtracts the reconstructed signal from the received SI \cite{chip}. This chip can suppress about 20 dB of the received SI power. Another technique, known as tapped delay lines (TDL), is proposed to predict the SI channels using attenuators and variable delays. However, such a circuit has to be properly tweaked to estimate the SI channel \cite{n17}. Even though TDL can suppress about 45 dB of the SI power, it is very complex to implement in multiple input multiple output (MIMO) FD systems since the analog circuit cannot properly scale with the increasing number of antennas. To overcome this problem, a spatial suppression technique has been proposed to mitigate the SI for MIMO FD system \cite{11}. Such a technique leverages the available degrees of freedom (DoFs) from the multiple antennas to cancel the SI, while maintaining an acceptable multiplexing gain. In particular, the analog beamforming technique requires only the implementation of the phase shifters, while the amplitudes have to be kept constant due to the huge power consumption of the analog-to-digital converter (ADC), digital-to-analog converter (DAC), phase shifters, and RF chain operating at mmWave frequencies. 

In this work, we consider a two-node FD system with multiple spatial streams and bidirectional links while assuming a hybrid architecture for each FD node. A modified version of the zero-forcing (ZF) max-power algorithm is proposed to design the beamformers to minimize the SI and maintain the losses imposed by the CA constraint at an acceptable level. This paper is organized as follows: Section II provides a description of the system and channel models. Fully-digital and hybrid beamforming designs are presented in Sections III and IV, respectively. Numerical results along with the discussion and analysis are detailed in Section V, while concluding remarks are given in Section VI.

\section{System Models}
\begin{figure}[!t]
    \centering
    \includegraphics[scale=.53]{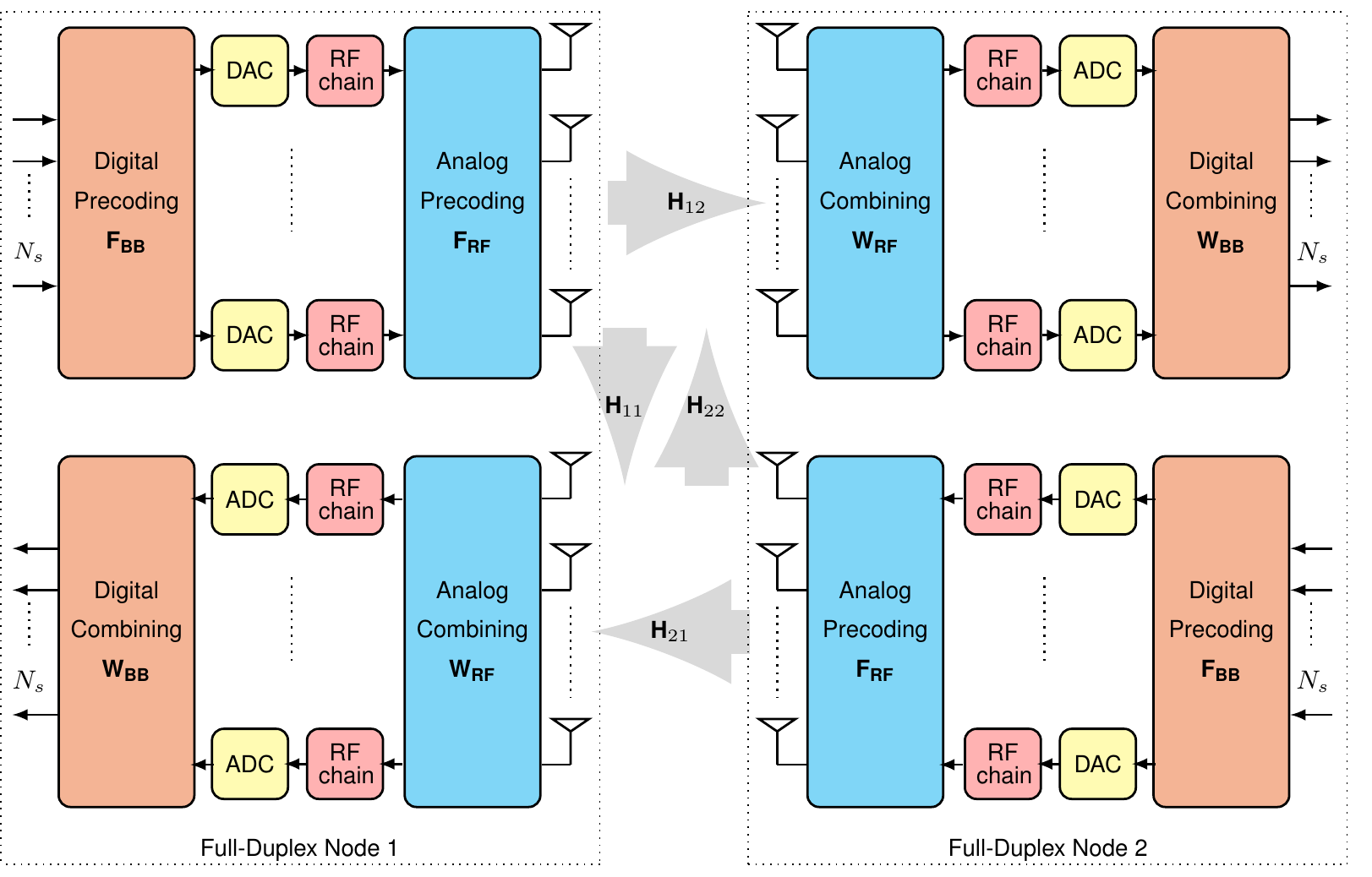}
    \caption{Architecture of a hybrid two-node FD network. }
    \label{system1}
\end{figure}

The architecture of the proposed hybrid two-node FD network is presented in Fig.~\ref{system1}, where each node is equipped with TX and RX antenna arrays, with multiple data streams supported in each direction. The combined received symbol at node $u$ transmitted from node $v$ is given by
\begin{equation}\label{eq1}
y_u = \sqrt{\rho_u} \textbf{W}_u^*\textbf{H}_{vu}\textbf{F}_{v}s_v + \sqrt{\tau_u}\textbf{W}^*_u \textbf{H}_{uu}\textbf{F}_{u}s_u + \textbf{W}^*_u\textbf{n}_u
\end{equation}
Here, $\textbf{W}_u$ and $\textbf{F}_v$ ($u,v \in \{1,2\}$) are the fully-digital combiner and precoder matrices of dimensions $N_{r,u}\times N_s$ and $N_{t,v}\times N_s$, respectively, with $N_{r,u}$ and $N_{t,v}$ being the number of RX and TX antennas at nodes $u$ and $v$, respectively, and $N_s$ being the number of spatial streams. In (\ref{eq1}), $\rho_u$ is the received power, $\tau_u$ is received SI power, and $\textbf{n}_u$ is the noise. The subscript $u$ denotes parameters at node $u$. Moreover, $\textbf{H}_{uv}$ is the channel from node $v$ to node $u$, $\textbf{H}_{uu}$ is the SI channel at node $u$, and $s_v$ is the transmit symbol at node $v$. Note that $(\cdot)^*$ is the Hermitian operator.
\subsection{Channel Model}
The stream from the TX node $u$ to the RX node $v$ is described by the channel matrix $\textbf{H}_{uv}$ with $u, v \in \{1,2\}$. To model channel $\textbf{H}_{uv}$, we adopt the geometric model based on the clusters and rays to capture the features of mmWave channel 
\begin{equation}{\label{{eq2.1}}}
\textbf{H}_{uv} = \sqrt{\frac{N_{t,u}N_{r,v}}{N_{\rm cl}N_{\rm ray}}} \sum_{k=1}^{N_{\rm{cl}}}\sum_{\ell=1}^{N_{\rm{ray}}}\alpha_{k,\ell}\textbf{a}_{\text{RX}}(\phi_{k,\ell}^v,\theta_{k,\ell}^v)  \textbf{a}^*_{\text{TX}}(\phi_{k,\ell}^u,\theta_{k,\ell}^u)  
\end{equation}
where $N_{\rm{cl}}$ and $N_{\rm{ray}}$ denote the number of clusters and rays, respectively, $\alpha_{k,\ell}$ is the complex gain of the $\ell^{\rm{th}}$ ray in the $k^{\rm{th}}$ cluster, and $\textbf{a}_{\rm t}$ and $\textbf{a}_{\rm r}$ are the antenna array steering and response vectors at TX and RX, respectively, evaluated at the angle of departure (AoD) $(\phi_{k,\ell}^u,\theta_{k,\ell}^u)$ at the TX node $u$ and at the angle of arrival (AoA) $(\phi_{k,\ell}^v,\theta_{k,\ell}^v)$ at the RX node $v$. In this channel model, we assume a uniform rectangular array (URA) of $N \times M$ dimensions where the array response is given by
\begin{equation}
\begin{split}
\textbf{a}(\phi,\theta) =& \frac{1}{\sqrt{NM}}\left[1,\ldots,e^{j\frac{2\pi}{\lambda}[pd_h\sin\phi\sin\theta+qd_v\cos\theta]},\ldots\right.\\&\left. ,e^{j\frac{2\pi}{\lambda}[(M-1)d_h\sin\phi\sin\theta+(N-1)d_v\cos\theta]}   \right]^{T}     
\end{split}
\end{equation}
where $\phi$ and $\theta$ are the azimuth and elevation angles, $M$ and $N$ are the horizontal and vertical dimensions, $\lambda$ is the signal wavelength, $d_h$ and $d_v$ are the antenna spacing in horizontal and vertical dimensions, respectively, $0 \leq p \leq M-1$ and $0 \leq q \leq N-1$ are the antenna indices in the 2D plane.

\subsection{Self-Interference Channel Model}
\begin{figure}[H]
    \centering
    \includegraphics{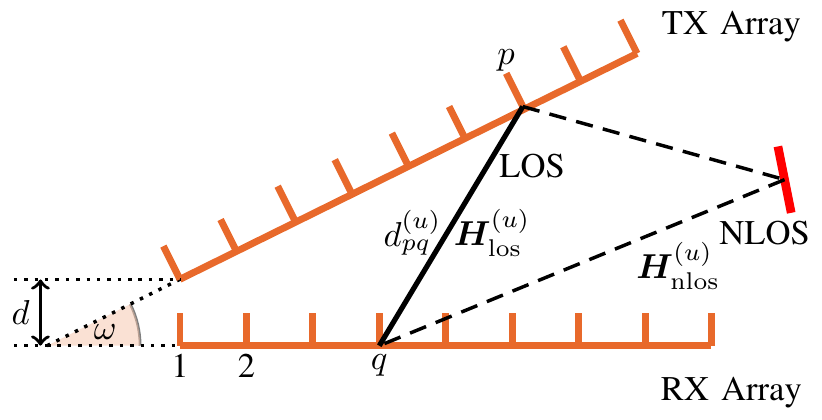}
    \caption{Relative position of TX and RX arrays at FD node $u$. Given that the TX and RX arrays are collocated, the far-field assumption that the signal impinges on the antenna
    array as a planar wave does not hold. Instead, for FD transceivers, it is more suitable to assume that the signal impinges on the array as a spherical wave for the near-field LOS channel.}
    \label{position}
\end{figure}

As illustrated in Fig.~$\ref{position}$, the SI leakage at node $u$ is modeled by the channel matrix $\textbf{H}_{uu}$, which is decomposed into line-of-sight (LOS) component modeled by $\textbf{H}^{(u)}_{\rm{los}}$ and non-line-of-sight (NLOS) leakage described by $\textbf{H}^{(u)}_{\rm{nlos}}$ which is a random complex Gaussian matrix. The LOS SI leakage matrix at node $u$ can be written as 
\begin{equation}\label{eq2.2}
 [\textbf{H}^{(u)}_{\rm{los}}]_{pq} = \frac{1}{d_{pq}^{(u)}}e^{-j2\pi\frac{d_{pq}^{(u)}}{\lambda}}    
\end{equation}
where $d_{pq}^{(u)}$ is the distance between the $p$-th antenna in the TX array and $q$-th antenna in the RX array at node $u$. The aggregate SI channel $\textbf{H}_{uu}$ can be obtained by
\begin{equation}\label{eq2.3}
\textbf{H}_{uu} = \sqrt{\frac{\kappa}{\kappa+1}}\textbf{H}^{(u)}_{\rm{los}} + \sqrt{\frac{1}{\kappa+1}}\textbf{H}^{(u)}_{\rm{nlos}}    
\end{equation}
where $\kappa$ is the Rician factor.
\section{Fully-Digital Beamforming Design}
Treating SI as noise, the sum rate of this network can be written as
\begin{equation}\label{eq4}
\begin{split}
\mathcal{I}_{\text{Digital}} &= \log\det\left( \textbf{I}_{N_s} + \frac{\rho_1}{N_s}\textbf{T}_1^{-1}(\textbf{W}_1^*\textbf{H}_{21}\textbf{F}_2)(\textbf{W}_1^*\textbf{H}_{21}\textbf{F}_2)^* \right)\\&+   \log\det\left( \textbf{I}_{N_s} + \frac{\rho_2}{N_s}\textbf{T}_2^{-1}(\textbf{W}_2^*\textbf{H}_{12}\textbf{F}_1)(\textbf{W}_2^*\textbf{H}_{12}\textbf{F}_1)^* \right) 
\end{split}
\end{equation}
where $\textbf{T}_u$ is the interference-plus-noise autocovariance matrix at node $u$ given by
\begin{equation}\label{cov}
\textbf{T}_u = \sigma_u^2\textbf{W}_u^*\textbf{W}_u + \tau_u(\textbf{W}_u^*\textbf{H}_{uu}\textbf{F}_{u}) (\textbf{W}_u^*\textbf{H}_{uu}\textbf{F}_{u})^*    
\end{equation}
where $\sigma_u^2$ is the noise variance given by
\begin{equation}
\sigma_u^2[{\rm{dBm}}] = -173.8 + 10 \log_{10}({\scriptsize{\textsf{Bandwidth}}}). 
\end{equation}
The design of the beamformers is based on the ZF-Max-Power to cancel the SI and maximize the rate ($\ref{eq4}$). The resulting problem is
\begin{equation}\label{eq6}
\mathcal{P}_1: \max\limits_{\textbf{F}_u,~\textbf{W}_u} {\mathcal{I}_{\text{Digital}}}
\end{equation}
\begin{equation}\label{eq7}
\text{s.t.}~~\|\textbf{F}_u\|^2_F =  \|\textbf{W}_u\|^2_F = N_s,~u\in \{1,~2\}
\end{equation}
\begin{equation}\label{eq8}
\textbf{W}_1^*\textbf{H}_{11}\textbf{F}_{1} = \textbf{W}_2^*\textbf{H}_{22}\textbf{F}_{2} = 0
\end{equation}
where $\|\cdot\|_F$ is the Frobenius norm. In the first step, we keep the digital precoders constant and solve for the digital combiners subject to the constraints ($\ref{eq7}$) and ($\ref{eq8}$). The maximization with respect to $\textbf{W}_1$ and $\textbf{W}_2$ is written as
\begin{equation}\label{eq9}
\max\limits_{\textbf{W}_1} \|\textbf{W}_1^*\textbf{H}_{21}\textbf{F}_2\|_F^2~~\text{s.t.}~~\|\textbf{W}_1\|^2_F = N_s,~\textbf{W}_1^*\textbf{H}_{11}\textbf{F}_{1} = 0  
\end{equation}
\begin{equation}\label{eq10}
\max\limits_{\textbf{W}_2}\|\textbf{W}_2^*\textbf{H}_{12}\textbf{F}_1\|_F^2~~\text{s.t.}~\|\textbf{W}_2\|^2_F = N_s,~\textbf{W}_2^*\textbf{H}_{22}\textbf{F}_{2} = 0.
\end{equation}
In the next step, we optimize the digital precoders while keeping the optimal digital combiners to their recent forms. This is translated into
\begin{equation}\label{eq11}
\max\limits_{\textbf{F}_1} \|\textbf{F}_1^*\textbf{H}_{12}^*\textbf{W}_2\|_F^2~~\text{s.t.}~~   \|\textbf{F}_1\|^2_F = N_s,~\textbf{F}_1^*\textbf{H}_{11}^*\textbf{W}_{1} = 0  
\end{equation}
\begin{equation}\label{eq12}
\max\limits_{\textbf{F}_2} \|\textbf{F}_2^*\textbf{H}_{21}^*\textbf{W}_1\|_F^2~~\text{s.t.}   ~~\|\textbf{F}_2\|^2_F = N_s,~\textbf{F}_2^*\textbf{H}_{22}^*\textbf{W}_{2} = 0.
\end{equation}
These four sub-problems share the same form and can be reformulated as
\begin{equation}\label{generic}
 \max\limits_{z} |z^*\beta|^2~~\text{s.t.}~~z^*\alpha = 0.
\end{equation}
Here, $z$ is a column vector of the precoder or combiner matrix, $\beta$ depends on the precoded or combined channel, and $\alpha$ defines the ZF null space.

The general form of the solution to the generic problem (\ref{generic}) is
\begin{equation}\label{sol}
    z = \left(\textbf{I} - \frac{\alpha \alpha^*}{\|\alpha\|^2}  \right)\beta
\end{equation}
where $\textbf{I}$ is an identity matrix. In the ZF cyclic maximization, vector $z$ is first solved using (\ref{sol}). It is then updated iteratively by replacing vector $\beta$ by the recent value of $z$. For the $n$-th iteration, vector $z$ is updated as
\begin{equation}
    z^{(n)} = \left(\textbf{I} - \frac{\alpha \alpha^*}{\|\alpha\|^2}  \right)z^{(n-1)}.
\end{equation}
Note that the design of the optimal digital beamformers is useful as a benchmarking tool to compare with the proposed hybrid beamforming design. In the next sections, we will consider the design of the analog and digital beamformers separately without the need for the optimal digital beamforming design obtained in this section.

\section{Proposed Hybrid Beamforming Design}
The above two nested cyclic optimizations yield the best column vector of the precoder and combiner matrices. Basically, such processes will be repeated for all the column vectors to design the precoder and combiner matrices.
\subsection{Analog Beamforming Design}
The analog beamforming design is based on maximizing the sum rate and cancelling the SI. In the analog domain, the sum rate (cost function) is given by

\begin{equation}\label{c_analog}
\begin{split}
\mathcal{I}_{\text{Analog}} =& \log\det\left( \textbf{I}_{N_{\text{RF}}} + \frac{\rho_1}{N_{\text{RF}}}\overline{\textbf{T}}_1^{-1} \overline{\textbf{H}}_{21} \overline{\textbf{H}}_{21}^* \right)\\&+   \log\det\left( \textbf{I}_{N_{\text{RF}}} + \frac{\rho_2}{N_{\text{RF}}}\overline{\textbf{T}}_2^{-1} \overline{\textbf{H}}_{12} \overline{\textbf{H}}_{12}^* \right) 
\end{split}
\end{equation}
where $N_{\text{RF}}$ is the number of RF chains, $\overline{\textbf{H}}_{vu} = \textbf{W}_{\text{RF},u}^* \textbf{H}_{vu} \textbf{F}_{\text{RF},v}$ is the effective channel and  $\overline{\textbf{T}}_u$ is given by
\begin{equation}
\overline{\textbf{T}}_u = \sigma_u^2 \textbf{W}_{\text{RF},u}^*\textbf{W}_{\text{RF},u} + \tau_u \overline{\textbf{H}}_{uu}\overline{\textbf{H}}_{uu}^*.   
\end{equation}
To maximize the cost function ($\ref{c_analog}$), the optimization problem is defined as follows
\begin{equation}\label{eq15}
\mathcal{P}_2: \max\limits_{\textbf{F}_{\text{RF},u},~\textbf{W}_{\text{RF},u}} \mathcal{I}_{\text{Analog}}
\end{equation}
\begin{equation}\label{eq16}
\text{s.t.}~~\|\textbf{F}_{\text{RF},u}\|^2_F =  \|\textbf{W}_{\text{RF},u}\|^2_F = N_{\text{RF}},~u\in \{1,~2\}
\end{equation}
\begin{equation}\label{eq17}
\textbf{W}_{\text{RF},1}^*\textbf{H}_{11}\textbf{F}_{\text{RF},1} = \textbf{W}_{\text{RF},2}^*\textbf{H}_{22}\textbf{F}_{\text{RF},2} = 0.
\end{equation}
Similar to the optimization problem constructed for the optimal digital beamformers, in the first step, we hold the analog precoders fixed and optimize the analog combiners.

\begin{equation}\label{eq18}
\max\limits_{\textbf{W}_{\text{RF},1}} \|\overline{\textbf{H}}_{21}\|_F^2~~\text{s.t.}~~\|\textbf{W}_{\text{RF},1}\|^2_F = N_{\text{RF}},~\overline{\textbf{H}}_{11} = 0  
\end{equation}

\begin{equation}\label{eq19}
\max\limits_{\textbf{W}_{\text{RF},2}} \|\overline{\textbf{H}}_{12}\|_F^2~~\text{s.t.}~~\|\textbf{W}_{\text{RF},2}\|^2_F = N_{\text{RF}},~\overline{\textbf{H}}_{22} = 0.  
\end{equation}
Then, we keep the recent forms of the analog combiners and optimize the analog precoders 
\begin{equation}\label{eq20}
\max\limits_{\textbf{F}_{\text{RF},1}} \|\overline{\textbf{H}}^*_{12}\|_F^2~~\text{s.t.}~~\|\textbf{F}_{\text{RF},1}\|^2_F = N_{\text{RF}},~~\overline{\textbf{H}}_{11}^* = 0 
\end{equation}
\begin{equation}\label{eq21}
\max\limits_{\textbf{F}_{\text{RF},2}}  \|\overline{\textbf{H}}^*_{21}\|_F^2~~\text{s.t.}~~\|\textbf{F}_{\text{RF},2}\|^2_F = N_{\text{RF}},~~\overline{\textbf{H}}_{22}^* = 0.  
\end{equation}
The analog beamformers must also satisfy the CA constraint, where each of the entries of the matrices $\textbf{F}_{\text{RF},1}$, $\textbf{F}_{\text{RF},2}$, $\textbf{W}_{\text{RF},1}$, and $\textbf{W}_{\text{RF},2}$ has a constant amplitude that equals to the inverse of the square root of the dimension of the corresponding matrix, such as
    \begin{equation}\label{eq22}
     [\textbf{F}_{\text{RF},u}]_{m,n} = \frac{[\textbf{F}_{\text{RF},u}]_{m,n}}{\sqrt{N_{t,u}N_{\text{RF}}}|[\textbf{F}_{\text{RF},u}]_{m,n}|}     
    \end{equation}
\begin{equation}
  [\textbf{W}_{\text{RF},u}]_{m,n} = \frac{[\textbf{W}_{\text{RF},u}]_{m,n}}{\sqrt{N_{r,u}N_{\text{RF}}}|[\textbf{W}_{\text{RF},u}]_{m,n}|}. 
\end{equation}
Thereby, each entry of the precoding and combining matrices has the following form
    \begin{equation}\label{eq23}
     [\textbf{F}_{\text{RF},u}]_{m,n} = \frac{e^{j\theta_{mn}}}{\sqrt{N_{t,u}N_{\text{RF}}}} 
    \end{equation}
    \begin{equation}
        [\textbf{W}_{\text{RF},u}]_{m,n} = \frac{e^{j\phi_{mn}}}{\sqrt{N_{r,u}N_{\text{RF}}}} 
    \end{equation}
where $\theta_{mn}$, and $\phi_{mn}$ are drawn from the feasible set of the phased shifters.

To fulfill the constraints, an outer iterative loop corresponding to the ZF cycle is performed to cancel the interference. For each outer iteration, an inner loop is conducted to maximize the sum rate (cost function in analog domain) and satisfy the CA constraint. These two nested loops should be run for each sub-problem ($\ref{eq18}$)-($\ref{eq21}$) to find the optimal analog precoders and combiners that maximize the rate, cancel the interference, and satisfy the CA constraint.

The sub-problems share the same generic form as 

\begin{equation}\label{sol2}
    z = \left(\textbf{I} - \frac{\alpha \alpha^*}{\|\alpha\|^2}  \right)\beta,~z \in \mathcal{C}^N
\end{equation}
where $\mathcal{C}^N$ is the CA subspace and $N$ is the subspace dimension, which in this case is the number of antennas at TX (if $z$ is a column vector of the precoder matrix) or RX (if $z$ is a column vector of the combiner matrix). The objective is to seek a vector $z$ that simultaneously maximizes the receiving power and minimizes the SI power. One way to solve such a problem is to apply the alternating projection method to find the solution vector $z$ \cite{16}. In particular, we will construct two nested optimization processes where the first cycle involves projection onto the ZF null space to satisfy the ZF constraint and the second cycle consists of projecting the vector onto the CA subspace $\mathcal{C}^N$. 

\subsection{Digital Beamforming Design}
The objective of designing the digital beamformers is to further maximize the sum rate (cost function) to compensate for the performance loss introduced by the CA constraint in the analog domain and manage the interference caused by the multiple spatial streams. Note that the SI cancellation takes place only in the analog domain. In the hybrid domain, the sum rate can be expressed as

\begin{equation}{\label{c_hybrid}}
\begin{split}
\mathcal{I}_{\text{Hybrid}} =&  \log\det\left(\textbf{I}_{N_s} + \frac{\rho_1}{N_s} \overline{\overline{\textbf{T}}}^{-1}_1 \overline{\overline{\textbf{H}}}_{21} \overline{\overline{\textbf{H}}}_{21}^*\right)  \\&+  \log\det\left(\textbf{I}_{N_s} + \frac{\rho_2}{N_s} \overline{\overline{\textbf{T}}}^{-1}_2 \overline{\overline{\textbf{H}}}_{12} \overline{\overline{\textbf{H}}}_{12}^*\right)   
\end{split}
\end{equation}
where $\overline{\overline{\textbf{H}}}_{vu} = \textbf{W}_{\text{BB},u}^* \overline{\textbf{H}}_{vu} \textbf{F}_{\text{BB},v}$, with $\textbf{W}_{\text{BB},u}$ being the digital combiner at node $u$ and $\textbf{F}_{\text{BB},v}$ being the digital precoder at node $v$, and $\overline{\overline{\textbf{T}}}_u$ is given by
\begin{equation}
\overline{\overline{\textbf{T}}}_u = \sigma_u^2 \textbf{W}_{\text{BB},u}^*\textbf{W}_{\text{RF},u}^*\textbf{W}_{\text{RF},u}\textbf{W}_{\text{BB},u} + \tau_u \overline{\overline{\textbf{H}}}_{uu}\overline{\overline{\textbf{H}}}_{uu}^*.   
\end{equation}
To maximize the cost function ($\ref{c_hybrid}$), the corresponding optimization problem is defined to find the best digital beamformers as  
\begin{equation}\label{eq26}
\mathcal{P}_3: \max\limits_{\textbf{F}_{\text{BB},u},~\textbf{W}_{\text{BB},u}} \mathcal{I}_{\text{Hybrid}}
\end{equation}
\begin{equation}\label{eq27}
\text{s.t.}~~\|\textbf{F}_{\text{BB},u}\|^2_F =  \|\textbf{W}_{\text{BB},u}\|^2_F= N_s.
\end{equation}
The optimization process consists of two steps. The first step is to keep the digital precoders fixed and optimize the digital combiners as
\begin{equation}\label{eq28}
\max\limits_{\textbf{W}_{\text{BB},1}} \|\overline{\overline{\textbf{H}}}_{21}\|_F^2~~\text{s.t.}   ~~\|\textbf{W}_{\text{BB},1}\|_F^2 = N_s
\end{equation}
\begin{equation}\label{eq29}
\max\limits_{\textbf{W}_{\text{BB},2}}\|\overline{\overline{\textbf{H}}}_{12}\|_F^2~~\text{s.t.}   ~~\|\textbf{W}_{\text{BB},2}\|_F^2 = N_s.
\end{equation}
This results in the optimal digital combiners given by
\begin{equation}\label{eq30}
\textbf{W}_{\text{BB},1} = 
\frac{\textbf{W}_{\text{RF},1}^*\textbf{H}_{21}\textbf{F}_{\text{RF},2}\textbf{F}_{\text{BB},2}}{\|\textbf{W}_{\text{RF},1}^*\textbf{H}_{21}\textbf{F}_{\text{RF},2}\textbf{F}_{\text{BB},2}\|_F}
\end{equation}
\begin{equation}\label{eq31}
\textbf{W}_{\text{BB},2} = 
\frac{\textbf{W}_{\text{RF},2}^*\textbf{H}_{12}\textbf{F}_{\text{RF},1}\textbf{F}_{\text{BB},1}}{\|\textbf{W}_{\text{RF},2}^*\textbf{H}_{12}\textbf{F}_{\text{RF},1}\textbf{F}_{\text{BB},1}\|_F}.
\end{equation}
The second step is to keep the optimal digital combiners found by ($\ref{eq30}$) and ($\ref{eq31}$) and optimize the digital precoders as

\begin{equation}\label{eq32}
\max\limits_{\textbf{F}_{\text{BB},2}} \|\overline{\overline{\textbf{H}}}_{21}^*\|_F^2~~\text{s.t.}   ~~\|\textbf{F}_{\text{BB},2}\|^2_F = N_s
\end{equation}
\begin{equation}\label{eq33}
\max\limits_{\textbf{F}_{\text{BB},1}} \|\overline{\overline{\textbf{H}}}_{12}^*\|_F^2~~\text{s.t.}   ~~\|\textbf{F}_{\text{BB},1}\|_F^2 = N_s.
\end{equation}
This results in the optimal digital precoders given by
\begin{equation}\label{eq34}
\textbf{F}_{\text{BB},1} = 
\frac{\textbf{F}_{\text{RF},1}^*\textbf{H}_{12}^*\textbf{W}_{\text{RF},2}\textbf{W}_{\text{BB},2}}{\|\textbf{F}_{\text{RF},1}^*\textbf{H}_{12}^*\textbf{W}_{\text{RF},2}\textbf{W}_{\text{BB},2}\|_F}
\end{equation}
\begin{equation}\label{eq35}
\textbf{F}_{\text{BB},2} = 
\frac{\textbf{F}_{\text{RF},2}^*\textbf{H}_{21}^*\textbf{W}_{\text{RF},1}\textbf{W}_{\text{BB},1}}{\|\textbf{F}_{\text{RF},2}^*\textbf{H}_{21}^*\textbf{W}_{\text{RF},1}\textbf{W}_{\text{BB},1}\|_F}.
\end{equation}
An iterative process is performed to update ($\ref{eq30}$), ($\ref{eq31}$), ($\ref{eq34}$), and ($\ref{eq35}$) until the cost function ($\ref{c_hybrid}$) is maximized or becomes constant.
\subsection{Upper Bound}
For the interference-free scenario, the optimal-capacity-achieving beamformers diagonalize the channel. By performing the singular value decomposition (SVD) of the matrix $\textbf{H}_{uv}$, we retrieve its singular values in descending order. This yields $N_s$ parallel sub-channels where the $n$-th sub-channel gain is given by the $n$-th singular value $\lambda_n(\textbf{H}_{uv})$. The upper bound of the sum rate is then given as

\begin{equation}\label{eq36}
\begin{split}
\mathcal{I}_{\text{Bound}} =& \sum_{n=1}^{N_s}\log\left( 1 + \frac{\rho_1}{N_s\sigma_1^2}\lambda_n(\textbf{H}_{21})^2 \right) \\&+ \sum_{n=1}^{N_s}\log\left( 1 + \frac{\rho_2}{N_s\sigma_2^2}\lambda_n(\textbf{H}_{12})^2 \right).
\end{split}
\end{equation}
It should be noted that such an upper bound cannot be achieved as the SI is not considered and a perfect feedback of the SVD precoders is assumed. However, it can be used as a tool to benchmark the proposed hybrid beamforming design and quantify the losses and limitations of the proposed technique.

\section{Numerical Results}
In this section, we discuss numerical results of the proposed system along with the analysis and comparisons with the existing work. For all cases, the simulations are validated with 1000 Monte Carlo iterations. First, we generate the channel realizations as well as SI channels. Next, we construct the precoders and combiners in the analog and digital domains for the hybrid architecture. For the analog-only architecture, we only need to design the analog beamformers and as a target performance, we can evaluate the statistics of the signal-to-interference-plus-noise-ratio (SINR) to get engineering insights into such a physical architecture. Finally, we evaluate the spectral efficiency with to various system parameters, which are summarized in TABLE \ref{param}.

\begin{figure}[!t]
\centering
\setlength\fheight{5.5cm}
\setlength\fwidth{7.5cm}
\input{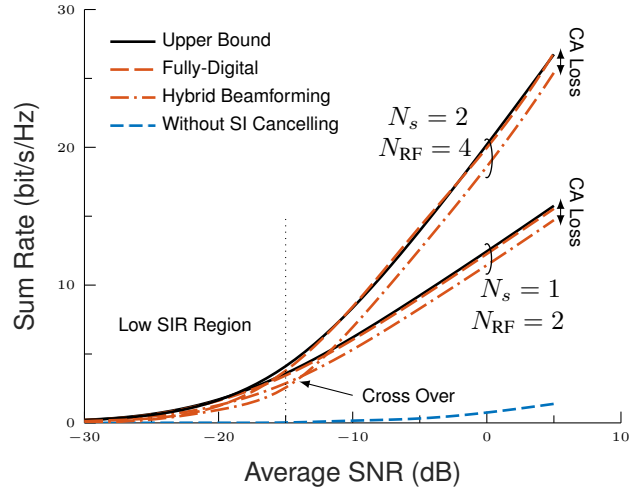}
    \caption{Sum rate performance of different systems. Comparisons are made between the upper bound, fully-digital, and hybrid beamforming designs, to illustrate the effects of the number of spatial streams and RF chains. Interference identified by the low signal-to-interference-ratio (SIR) region are caused by the multiple spatial streams. The short-dashed blue curve is obtained by SVD precoding and MMSE combining. }
    \label{pict1}
\end{figure}

Fig.~\ref{pict1} illustrates the performance of the proposed system at different settings. We clearly observe that the performance is substantially improved with the increase of the spatial streams and the number of RF chains. This is explained by the fact that these parameters improve the system resolution and increase the number DoFs. The DoFs are mainly observed at high signal-to-noise-ratio (SNR) where the configuration $N_s = 2, N_{\text{RF}} = 4$ achieves better multiplexing gain compared to the configuration $N_s = 1, N_{\text{RF}} = 2$. At the low SNR, we observe that the system achieves relatively better results for lower number of spatial streams. In fact, the performance is affected by the interference caused by the multiple streams and hence the per-stream-rate is degraded at low power regime where the interference dominates. As the SNR increases, the signal power increases and dominates the multiple streams interference whose effect is mitigated by digital precoding and combining. Moreover, the loss incurred by the CA constraint is observed to be around 1 to 2 bps/Hz and this gap seems to be irreducible and independent to the system settings (this gap is the same for the two configurations). In addition, the performance for SVD precoding and minimum mean squared error (MMSE) combining is highly degraded by the SI compared to the proposed system. In fact, this method completely ignores the interference, which results in beamformers that are not necessarily orthogonal to the ZF subspace. Consequently, the system is highly affected by the SI, leading to a poor spectral efficiency.

\begin{figure}[H]
\centering
\setlength\fheight{5.5cm}
\setlength\fwidth{7.5cm}
\input{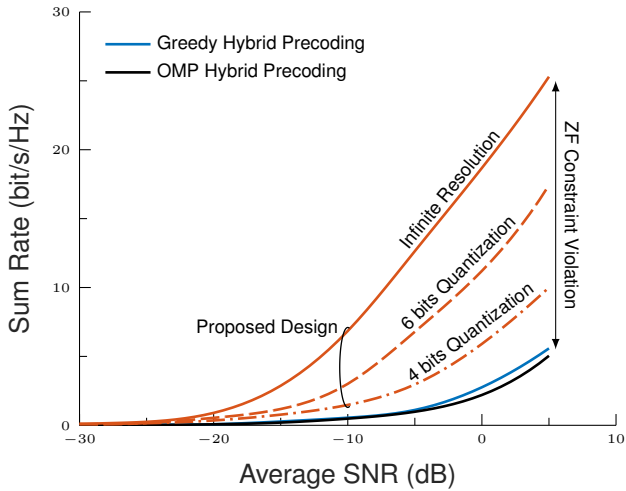}
    \caption{Comparison of the proposed hybrid beamforming design with greedy and OMP hybrid precoding schemes. We consider the proposed design without and with quantization of 6 and 4 bits. }
    \label{pict2}
\end{figure}

Fig. \ref{pict2} illustrates the performance comparison between the proposed design and the orthogonal matching pursuit (OMP) and greedy hybrid precoding schemes. For the latter two schemes, we perform the fully-digital beamformers detailed in Section II and split the fully-digital solution into analog and digital beamformers. It is clear from the figure that the performances achieved by these schemes are poor since the decomposition violates the ZF constraint which induces severe loss to the spectral efficiency. The impact of the phase quantization on the system is also considered. From Fig.~\ref{pict2}, the performance improves with higher order quantization.
\begin{table}[!t]
\renewcommand{\arraystretch}{1}
\setlength{\arrayrulewidth}{.7pt}
\caption{Simulation Parameters}
\label{param}
\centering
\begin{tabular}{|l|c|l|c|}
\hline\hline
Carrier frequency& 28 GHz&Bandwidth & 850 MHz\\
\hline
TX antennas& 16&RX antennas& 16\\
\hline
Number of clusters& 6&Number of rays& 8\\
\hline
Angular spread& 20$^{\circ}$ & Transceivers gap ($d$) & 2$\lambda$\\
\hline
Transceivers incline ($\omega$)& $\frac{\pi}{6}$ & Rician factor & 5 dB\\
\hline
SI power ($\tau$)& 30 dB & Antenna separation & $\frac{\lambda}{2}$\\
\hline
Spatial streams ($N_s$)& 2 & RF chains ($N_{\text{RF}}$) & 4\\
\hline\hline
\end{tabular}
\end{table}

\begin{figure}[H]
\centering 
\setlength\fheight{5.5cm}
\setlength\fwidth{7.5cm}
\input{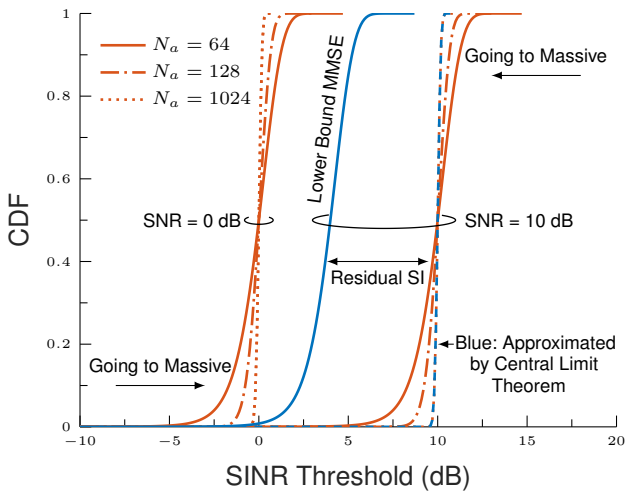}
    \caption{Performance comparison between analog-only beamforming schemes. CDF of the SINR for different array configurations at two values of the average SNR. In this simulation, we assume the same number of TX and RX antennas $N_a$. The lower bound MMSE performance is simulated with $N_a = 64$ \cite{unconst}.}
    \label{pict3}
\end{figure}

Fig. \ref{pict3} provides the cumulative distribution function (CDF) of the SINR for different numbers of antennas and for two SNR values (0 and 10 dB). We observe that the performance of the proposed system abruptly changes and saturates around 0 and 10 dB of SINR which exactly corresponds to the SNR values considered in this scenario. This result is very important as the receiving SNR coincides with the SINR and this shows that the proposed design completely eliminates the interference, thanks to the ZF constraint. The beamforming design known as the lower bound MMSE \cite{unconst}  is reproduced at an SNR of 10 dB. It is observed that at this SNR value, the lower bound MMSE performance is shifted to the left side relative to the proposed design and it saturates at around 4 dB. This shows that the design in \cite{unconst} achieves around 24 dB of SI reduction while a residual SI of 6 dB margin is left, which limits the system performance. 

\section{Conclusion}
In this paper, we proposed a modified version of the ZF max-power algorithm to design the hybrid beamformers for a bidirectional two-node FD system. We showed that the proposed design completely eliminates the SI power thanks to the ZF constraint, which outperforms the lower bound MMSE, OMP, and greedy hybrid precoding. Unlike the previous works, our algorithm provides an efficient beamforming design that minimizes the rate losses due to the CA constraint using the alternating projection method. Although the FD beamforming design is very challenging for wideband systems, such models deserve careful investigation for future 5G research as mmWave channels are generally frequency-selective.

\bibliographystyle{IEEEtran}
\bibliography{main.bib}
\end{document}